# Cloudless-Training: A Framework to Improve Efficiency of Geo-Distributed ML Training

Wenting Tan[1,2], Xiao Shi[1,*], Cunchi Lv[1,2], Xiaofang Zhao[1]

[1] Institute of Computing Technology, Chinese Academy of Sciences, Beijing, China
[2] University of Chinese Academy of Sciences, Beijing, China
{tanwenting, shixiao, lvcunchi21s, zhaoxf}@ict.ac.cn

*Abstract*—Geo-distributed ML training can benefit many emerging ML scenarios (e.g., large model training, federated learning) with multi-regional cloud resources and wide area network. However, its efficiency is limited due to 2 challenges. First, efficient elastic scheduling of multi-regional cloud resources is usually missing, affecting resource utilization and performance of training. Second, training communication on WAN is still the main overhead, easily subjected to low bandwidth and high fluctuations of WAN.

In this paper, we propose a framework, Cloudless-Training, to realize efficient PS-based geo-distributed ML training in 3 aspects. First, it uses a two-layer architecture with control and physical training planes to support elastic scheduling and communication for multi-regional clouds in a serverless maner. Second, it provides an elastic scheduling strategy that can deploy training workflows adaptively according to the heterogeneity of available cloud resources and distribution of pre-existing training datasets. Third, it provides 2 new synchronization strategies for training partitions among clouds, including asynchronous SGD with gradient accumulation (ASGD-GA) and inter-PS model averaging (MA). It is implemented with OpenFaaS and evaluated on Tencent Cloud. Experiments show that Cloudless-Training can support general ML training in a geo-distributed way, greatly improve resource utilization (e.g., 9.2%-24.0% training cost reduction) and synchronization efficiency (e.g., 1.7x training speedup over baseline at most) with model correctness guarantees.

*Keywords—geo-distributed ML training, serverless, elastic scheduling, synchronization*

## I. Introduction

Geo-distributed ML training can satisfy the requirements of large-scale or collaborative training with multi-regional flexible cloud resourcing and partitioned training dataset due to business or privacy principles. For example, the real-time available resources in a single cloud may fail to satisfy the training of a large model, and multimodal training datasets (e.g., user behavior logs, images, videos) can be produced at extremely high rates and stored in cloud around the world. All these drive the necessity of geo-distributed ML training on multi-regional clouds, as shown in Fig. 1. Currently, it is used in many emerging ML scenarios, such as large model training[1], federated learning[2], and ML training in collaborative edge-cloud[3] or mixed cloud[4] environments.

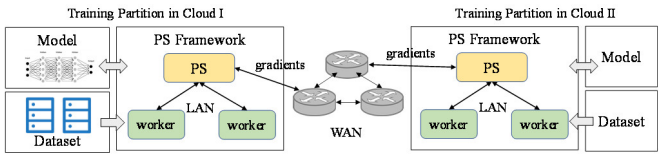

Fig. 1. The architecture of geo-distributed ML training method

However, it is non-trivial to acquire efficient geo-distributed training for users due to challenges in managing load balancing and communication in the complex environment. First, efficient elastic scheduling of multi-regional cloud resources is usually missing, resulting in load imbalance. It decreases resource utilization, speed, and performance of training directly. For example, training LeNet[43] in 2 cloud regions of Tencent Cloud with the same CPU cores and uneven data distribution leads to 25% resource over-provisioning in 1 region, as shown in Fig. 2. Second, communication on WAN for model synchronization is still the main overhead, easily subjected to low bandwidth and high fluctuations of WAN. To keep the model state consistent among clouds, communication for model synchronization is required to periodically exchange model replicas of each training partition. With Parameter Server (PS) pattern[12], each cloud-level PS needs to communicate and update the global model replica, the communication overhead grows depending on the model size and network bandwidth, which is quite stressed for the WAN among clouds. For example, the WAN communication time in training ResNet18[6] with GPU in 2 cloud regions of Tencent Cloud with 100Mbps WAN takes up over 98% of total time, as shown in Fig. 3.

Studies have been explored to support various geo-distributed ML scenarios, including both training and inference[8][9][2][3][4]. Gaia decouples the communication within a cloud from the communication between clouds and presents a new ML synchronization model, Approximate Synchronous Parallel (ASP) for dynamically eliminating insignificant communication between clouds[8]. Nebula-I adopts hybrid parallel computing methods and adaptive communication acceleration techniques with data compression methods[9]. They propose methods to improve synchronization efficiency on WAN, but not the load balancing challenge mentioned above. For communication optimization, their models are heuristic and can be further studied.

In this paper, we propose Cloudless-Training, a framework that improves resourcing and synchronization efficiency of geo-distributed ML training from 3 aspects. First, it uses a two-layer architecture to decouple controlling and physical training with serverless paradigm, simplifying the logical view of training management with multi-regional elastic scheduling and synchronization support. Second, it provides an elastic scheduling strategy that deploys training workflows with proper resource allocations, according to the heterogeneity of available cloud resources and distribution of pre-existing training datasets among clouds. Third, it introduces 2 model synchronization strategies for training partitions among clouds, including asynchronous SGD with gradient accumulation (ASGD-GA) and model averaging (MA). They are variants and deliberately regulated to match geo-distributed requirements. It is implemented based on

OpenFaaS[7] and evaluated in 2 regions (Shanghai and Chongqing) of Tencent Cloud with 100Mbps WAN.

In summary, we make 4 major contributions:

1. We design two layered PS-based geo-distributed ML training framework with serverless paradigm, greatly improving resource utilization and synchronization efficiency in training.

2. We provide an elastic scheduling strategy that guarantees load-balanced resource provisioning by modeling heterogeneity of available cloud resources and distribution of pre-existing training datasets.

3. We adopt 2 kinds of strategies for multi-regional model synchronization among training partitions, including asynchronous SGD with gradient accumulation (ASGD-GA) and model-averaged (MA). Both can significantly reduce the overhead over WANs.

4. We implement and evaluate Cloudless-Training in a real multi-regional cloud environment, showing that resourcing utilization and training performance of geo-distributed training can be greatly improved. It can improve resource utilization (e.g.,9.2%-24.0% training cost reduction), and synchronization efficiency (e.g., 1.7x speedup of training over baseline at most) with model correctness guarantees.

## II. BACKGROUND AND CHALLENGES

### A. Geo-Distributed ML Training

Compared with trivial ML scenarios, many emerging ML applications bring new requirements and considerably drive the practice of geo-distributed ML training, to realize high-performance, large-scale, or collaborative training. Here, we conclude 2 basic requirements.

**Requirement 1. To flexibly acquire more cloud resources than in a single cloud.** With the ever-growing of model size and the volume of datasets, the real-time available resources in a single cloud may fail to satisfy the training of a large model. AI applications keep asking for computing resources with both higher quantity and quality[17][18]. For example, Megatron LM model developed by Nvidia has 8.3 billion parameters and is trained on 512 Nvidia Tesla V100s for 174 GB text as corpus[29]. This makes that small-scale cloud data centers are unlikely to satisfy their requirements sometimes. It is natural to acquire supplements from other clouds by splitting training tasks into multiple clouds. Based on the strategy, geo-distributed ML training can be used to support training with various models, in edge-cloud[19], mixed-cloud[4] or multi-cloud environments.

**Requirement 2. To utilize data distributed across multi-regional clouds for training.** Concerning latency or QoS, many applications are distributed across multi-regional clouds. The data is usually stored locally and is valuable for building AI-assistant services. It needs deliberate consideration of how to realize ML training in this situation. The data (e.g., user behavior logs, images, videos) can be produced at extremely high rates around the world [20][21]. For example, TikTok users produce massive information by watching videos, served by its globally deployed cloud services[22]. Gathering all data together faces practical challenges, like low bandwidth and high-cost transportation on WAN, and strict data privacy rules, for which federated learning[2] serves. Thus, geo-distributed ML training can be considered to coordinate training across multiple clouds.

**Goals.** To satisfy the requirements, it is attractive to build a highly efficient geo-distributed ML training framework, which is scalable and heuristic for further practices and studies. Commonly seen parallel training patterns include data parallelism [14], model parallelism[15] and pipeline parallelism[16]. In this work, we focus on improving data parallelism training with PS architecture[12], which classically supports data parallelism. The working efficiency of geo-distributed ML training with PS is still limited. Although studies have proposed many solutions[8][9], the frameworks they build take no account of the rising cloud computing trend, like serverless computing[24], which is helpful to reduce management burdens in training. In this work, we argue that serverless can ease the framework building and help better satisfy the requirements due to its essential advantage of scaling out with fine-grained cloud functions. Besides, we observe that the load imbalance and communication overhead are 2 main factors, resulting in prominent challenges in practicing geo-distributed ML training.

### B. Challenge 1: Elastic Scheduling across Clouds

In geo-distributed ML training, load imbalance happens easily across clouds, asking for efficient elastic scheduling.

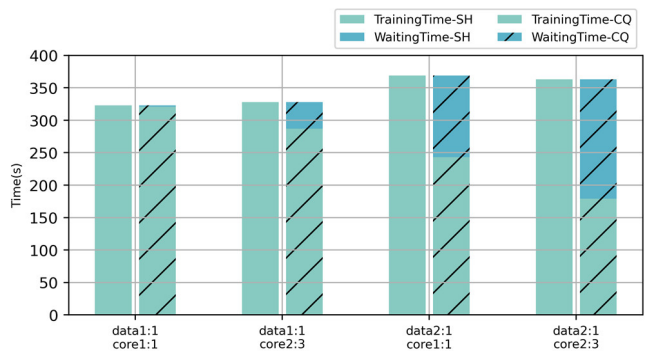

Fig. 2. The time proportion of training LeNet with various kinds of heterogeneous resource allocations and uneven data distributions in Shanghai and Chongqing regions of Tencent Cloud

**Load imbalance across multi-regional clouds.** It is usually empirical or rough (e.g., with greedy strategy) to set resourcing plans for training tasks in each cloud. Considering the condition of heterogeneous resources and uneven data distribution in multi-regional clouds, it is difficult to get a proper setting to avoid load imbalance. Then the problem can decrease resource utilization, speed, and performance of training explicitly. As shown in Fig. 2, as the data distribution ratio and the type and amount of available computing resources change, load imbalance is generated, and the training speed is totally decided by the peer with the most load. Meanwhile, peers with less load will hold computing resources while waiting for straggler peers to catch up, causing unnecessary resource consumption. Meanwhile, this can amplify the degree of training asynchronization in each cloud and decreases the performance explicitly as peers may receive heavily stale gradients or parameters from stragglers. For example, the AdamLike model[25] shows that gradient staleness severely hinders the convergence process.

**Goals.** To achieve high resource utilization, speed, and performance in geo-distributed ML training, it is necessary to balance the relative load among clouds, so that training in each cloud presents a more coordinating pattern to process However, former studies usually assume that training is carrying out in a load balancing way with homogeneous resources and event data distribution among clouds, such as Gaia[8], Nebula-I[9].To make load balancing scheduling dynamically match each training, we build a model to quantify load condition of available resources and the data distribution. Besides, we promote to use serverless to improve resource scaling efficiency by allocating and recycling training workers on demand, reducing long-term waiting for stragglers.

*C. Challenge 2: Reducing Synchronization Overhead on WAN*

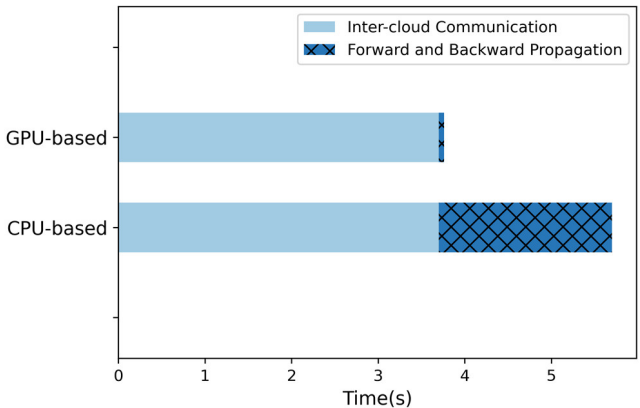

Fig. 3. The time proportion of training ResNet18 with CPU or GPU in 2 Shanghai and Chongqing regions of Tencent Cloud

In geo-distributed ML training, synchronization of intermediate results of models on WAN leads to much more overhead than on LAN, affecting (e.g., pausing, delaying) training processes periodically in running time. The overhead is nonnegligible.

**Synchronization overhead on WAN.** In ML training, each worker needs to communicate frequently with PS or other workers to keep the model updated. However, in geo-distributed ML training, part of communication converts to an inter-cloud pattern through WAN, which usually can only provide bandwidth with hundreds of Mbps (e.g., 100Mbps), at least 50 times slower than LAN in the cloud[23]. This increases the communication proportion in training. As Fig. 3 shows, while training ResNet18[6] (model size 48MB) on Chongqing and Shanghai regions of Tencent cloud, the communication time on WAN takes up over 64.9%, and 98.4% of total training time with CPU and GPU respectively. It not only slows down the training process but also may bring more cost for users due to charging for WAN traffic or indirect resourcing waste from waiting for synchronization.

**Goals.** It is helpful to reduce the synchronization overhead on WAN for better training speed, performance, and lower cost. In studies to optimize synchronization in trivial ML training, there are mainly 2 ways: compressing data size of synchronization, like DGC[13], top-K[35], or reducing synchronization frequency, like gradient accumulation[26], model averaging[27]. Gradient accumulation strategies accumulate the local gradients until it meets the condition or reaches the threshold for synchronization. Model averaging strategies periodically average models individually trained by parallel workers.

Work like Gaia[8] adopts synchronization frequency reduction, decoupling intra-cloud communication from inter-cloud communication, and using Approximate Synchronous Parallel (ASP) to eliminate insignificant communication. However, the significance threshold for updating model is hard to define accurately because tiny parameter updates may affect model convergence. In this paper, the communication addressing mechanism is considered first to serve cloud function communication the serverless architecture of Cloudless-Training. On the other hand, 2 strategies, variants of former-mentioned methods, are adopted with efforts of synchronization frequency reduction, including asynchronous SGD with gradient accumulation (ASGD-GA) and inter-PS model averaging (MA). As far as we know, they are adopted in PS-based geo-distributed ML training for the first time. ASGD-GA provides an asynchronous coordination policy for training partitions, which is usually synchronous in former work, and integrates gradient accumulation to reduce information loss, while MA is a model parameter oriented method.

I. CLOUDLESS-TRAINING DESIGN

*A. Framework Overview*

**Architecture.** Cloudless-Training provides a holistic view for realizing PS-based geo-distributed ML training. It is structured in 2 layers, including a control plane and a physical training plane. It provides a logical view that all training partitions are gathered into a unified logical plane from different clouds, as shown in Fig. 4. Compared to traditional geo-distributed ML training framework, Cloudless-Training is built in a serverless style. In the control plane, it is mainly in charge of scheduling training tasks and supporting communication addressing among all cloud members, so that each cloud-level training partition can work in parallel and acquire global communication addresses of peers' PS communicator cloud functions. The control plane can be deployed in any clouds via serverless, so that users can submit training tasks through it by providing model definitions and training configurations. In the physical training plane, training in each cloud, forming a partition, and is organized, deployed, and running as a serverless workflow. Internally, the workflow completes the functionalities of trivial ML training. Externally, the workflow periodically communicates with others via a communicator function, updates local model state for local use. Except for in the startup phase of training, all training partitions in each cloud work in a decentralized way.

**Training workflow adjustment.** According to the architecture, the complete training workflow works across multiple clouds. Compared to the trivial training serverless workflow, the training workflow in Cloudless-Training makes 2 main adjustments. First, it inserts cloud functions in the control plane for coordination, including a scheduling function and a communicator addressing function. They work at the startup phase. Second, it modifies functionality of PS cloud functions for synchronization on WAN. Then the modified PS functions in each cloud can expose its identity on WAN through communicator functions and manage the intermediate state with other PS functions.

The workflow scales in serverless manners to serve a training. When a training request arrives, the scheduler function responds first, loads the scheduling strategy, generates training plans for each cloud, and invocates sub workflows in each cloud. Then, the global communicator function waits for PS function in each cloud to be ready, and assigns communication addresses for each PS communicator mapping their serverless identities with <IP, Port> on WAN.

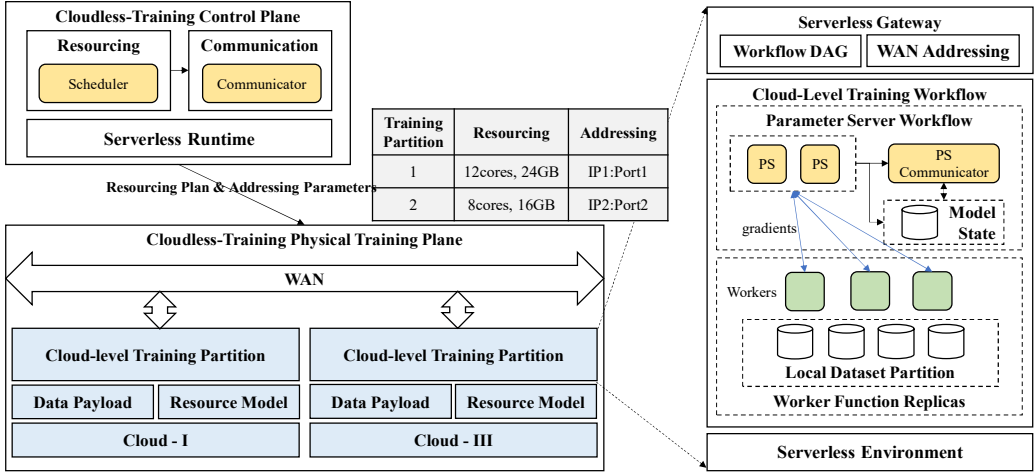

Fig. 4. Architecture of Cloudless-Training

While the preparation is done, each cloud-level training partition deploys corresponding training functions and executes them locally, including accessing local training data, computing local stochastic gradients, and updating local model descriptions. At predetermined time points (e.g., every 10 iterations) for synchronization, each PS communicator transits the local model state to another PS. When a PS receives a remote model state, it updates its local state from it, and lets local workers access the new version after. To reduce resource consumption, worker functions in each cloud are terminated immediately after the local training finishes (e.g., when epoch count or loss reaches the threshold locally).

**Scheduling mechanism.** The scheduling mechanism can support centralized scheduling strategies and be extended for more strategies, if necessary, by updating the implementation of the scheduler cloud function. In this work, we propose an elastic scheduling strategy to achieve load-balanced resource provisioning among clouds, as discussed in section B. The mechanism allows to probe or manually set available cloud resources and data distribution status to evaluate the heterogeneity of clouds and generate resourcing plans according to the model of elastic scheduling strategy.

**Synchronization support.** In trivial ML training, since cloud functions are elastic and the PS cloud functions are stateful, each PS has an address in its local serverless runtime and can be consistently accessed by local functions. While for WAN communication, the global communicator function needs to assign a unique identity for each PS communicator for WAN communication. To cut communication traffic on WAN, Cloudless-Training limits each PS to send its state to only one other PS each time. Thus, the communicator needs to plan the communication topology and notify each PS in preparation or when rescheduling happens. Furthermore, it provides 2 kinds of strategies that can reduce synchronization frequency across multiple clouds with training correctness guarantees, including asynchronous SGD with gradient accumulation and inter-PS model-averaged method, as discussed in section C.

*B. Elastic Scheduling Strategy*

The elastic scheduling strategy is inspired by the time composition of various training phases. It quantifies the relative computation load to balance the time consumption of the key computation phase, and then forms a heuristic scheduling algorithm.

**Scheduling insight and load modeling.** Divided by the synchronization moment on WAN each time, the intervals can be seen as a repeating period ($T_{process}$) of local processing in each training partition. In the interval, each cloud-level training completes model loading ($T_{load}$), local training ($T_{train}$) with forward and backward propagation and local communication and waits for peers in other clouds to catch up or finish. Since the training process is stateful and cloud resources will not be released while training, the waiting time causes unnecessary resource consuming. The relationship can be represented as $T_{process} = T_{load} + T_{train}$. As discussed in section II.B, if $T_{process}$ of each cloud differs largely, the performance of training cannot be guaranteed. According to studies[11] and common sense, the $T_{train}$ is the main part of $T_{process}$, and can be directly affected by the load balance status.

TABLE I. Training speed quantification of cloud resources

| CPU/GPU | Used Cores | TFLOPS/ TN | Iteration Time (s)/IN | IN/TN ratio |
|---|---|---|---|---|
| Intel Xeon IceLake (baseline) | 2 | 0.096/ 1.000 | 3.697/ 1.000 | 1.000 |
| Intel Xeon Cascade Lake | 2 | 0.090/ 0.938 | 5.549/ 0.666 | 0.710 |
| Intel Xeon Skylake | 2 | 0.112/ 1.167 | 3.800/ 0.973 | 0.834 |
| Nvidia T4 | 2560 | 5.554/ 57.854 | 0.062/ 59.629 | 1.031 |
| Nvidia V100 | 5120 | 13.345/ 139.010 | 0.024/ 154.042 | 1.108 |

The load balance status can be roughly predicted by considering the matching of allocated cloud resources and data distribution. By sampling some computing devices in clouds, including both CPUs and GPUs, we collect their specifications (e.g., type, number of cores of CPU or CUDA, TFLOPS ), quantify their computing power by observing practical performance (e.g., iteration time of training Resnet18 with dataset cifar-10[28], and normalizes metrics with baseline (e.g., TFLOPS normalization(TN), iteration time normalization (IN), and the ratio of IN/TN) for comparisons in training, as shown in TABLE I. It is obvious that the value of $T_{train}$ (quantified by iteration time) is inversely optional to the computing power of devices ($C_{device}$, quantified by TN) in each training iteration. In total, $T_{train}$ is also directly proportional to the size of dataset ($S_{data}$). Thus, we get $T_{train} \propto \frac{S_{data}}{C_{device}}$. For each device, $C_{device}$ can be roughly predicted or compared by its TFLOPS value or empirical iteration time.

Then, the training load power (**LP**) of a specific cloud i can be defined. It describes the capacity to handle a specific load with specific resources. It can be roughly quantized as the sum of the computing power of available devices (e.g., CPU, GPU) to the size of local datasets, as shown in (1). Here, m, n represents the amount of CPU, and GPU devices that can be allocated for the training, which may be reserved or applied in real-time by users. This can be used to predict relative load status of training with various resource allocation schemes, which is an important indicator in the scheduling algorithm.

$$LP_i = \frac{\sum_{m=1}^{M} N_{cpu,m} * P_m + \sum_{n=1}^{N} N_{gpu,n} * P_n}{S_{data}} \quad (1)$$

**Scheduling algorithm.** Furthermore, we propose an algorithm to plan training resourcing for each cloud to get load-balanced elastic scheduling, as shown in TABLE II. The key idea is to compare the load power unit of each cloud ($LP_i$), find the smallest load unit, which can be the worst straggler, and use it as a reference substance to search the practical number of resources for each cloud to match the straggler (search_optimal_plan()) by brutal force.

TABLE II. Elastic Scheduling Algorithm

| **Algorithm 1** Optimal Matching Algorithm | |
|---|---|
| 1 | **INPUT:** |
| 2 | N: Count of clouds, |
| 3 | Res[N]: Computing resources of all partitions, |
| 4 | $S_{data}$[N]: Dataset size of all partitions. |
| 5 | **OUTPUT:** |
| 6 | ResPlan[N]: Resourcing plan for each cloud. |
| 7 | |
| 8 | LP[N] = {} |
| 9 | MinLP = +∞ |
| 10 | ResPlan[N] = {} |
| 11 | **for** i, res **in** Res[N] **do** |
| 12 | LP[i] ← get_LP(res)    // formula (1) |
| 13 | MinLP ← **min**(MinLP, LP[i]) |
| 14 | **end for** |
| 15 | **for** i, res **in** Res[N] **do** |
| 16 | ResPlan[i] ← search_optimal_plan(res) |
| 17 | **end for** |
| 18 | **return** ResPlan[N] |

### C. Optimization of Synchronization on WAN

Cloudless-Training builds a basic WAN synchronization mechanism among workflows based on the WAN communication addressing support, and provides 2 strategies, asynchronous SGD with gradient accumulation (ASGD-GA) and inter-PS model averaging method (MA). They support the synchronization of gradients and model parameters respectively by reducing synchronization frequency without hurting model correctness. ASGD-GA provides an asynchronous coordination policy for geo-distributed training partitions, which is usually synchronous before, and integrates gradient accumulation to reduce information loss. Differently, MA is mainly model parameter oriented. Thus, these 2 strategies can be representative of gradient-based and parameter-based model synchronization methods.

The basic WAN synchronization mechanism works in a periodical way to synchronize the model state between sender and receiver PS communicators among clouds as follows:

1. In each training iteration, worker functions of each training partition pull the latest model from the local PS, execute SGD, and push gradients to the local PS.
2. The local PS updates the local model state with the SGD. The local update process can be synchronous or asynchronous.
3. The local PS checks the synchronization condition of synchronization strategy (ASGD-GA or MA) to decide whether model synchronization with other clouds is needed now. If not, the local PS finishes the process.
4. The local PS communicator packs the state to be sent and communicate with the receiver PS according to the settings of synchronization strategy (ASGD-GA or MA).
5. The receiver PS updates its local model state according to the algorithm of synchronization strategy.

To be more specific, the settings of synchronization condition, state to be sent (gradient or model parameters), communication pattern (synchronous or asynchronous among all training partitions), update algorithm (SGD or model averaging) vary in ASGD-GA and MA.

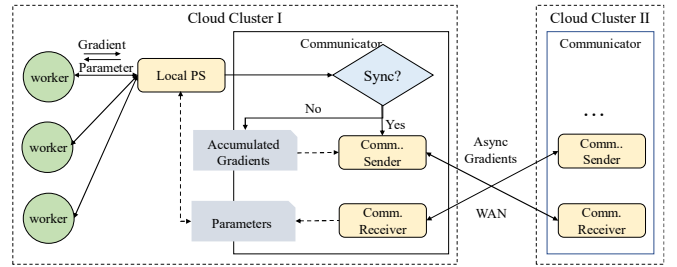

Fig. 5. ASGD-GA synchronization process

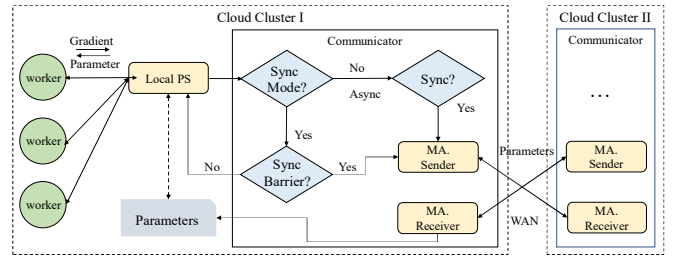

Fig. 6. Inter-PS MA synchronization process

**Asynchronous SGD with gradient accumulation (ASGD-GA).** ASGD-GA focuses on synchronizing gradients, as shown in Fig. 5. Its synchronization condition is defined as a synchronization frequency variable. For the state to be sent, it uses the local accumulated gradient. While the synchronization condition is not reached, the gradients will be kept accumulating by merging newly generated gradients. The communication pattern uses an asynchronous way so that each training partition will not be blocked to wait for other partitions to be ready for model synchronization. For model update of receiver PS, it uses SGD method.

**Inter-PS Model Averaging (MA)**. In distributed ML systems, model averaging is a periodic synchronization method that can cut down communication traffic between workers. In Cloudless-Training, it is used for cutting down communication traffic between PS among clouds, as shown in Fig. 6. MA focuses on improving the experience of handling model parameters. For the sending pattern, it can be synchronous or asynchronous. For the model state to be sent, it uses local model parameters. For the synchronization condition, it uses synchronization barrier in a synchronous communication pattern, and uses synchronization frequency variable in the asynchronous communication pattern.

## II. IMPLEMENTATION

The implementation of Cloudless-Training adopts OpenFaaS[7] as a basic serverless runtime framework. The development can be divided into 2 parts, including OpenFaaS customization for multi-regional cloud computing and workflow adjustment.

For the OpenFaaS customization, we make 2 extensions. First, OpenFaaS is enhanced to support serverless workflow deployment and invocation. Workflow is added as a new entity in OpenFaas, allowing to define DAG of workflow. The OpenFaaS gateway is extended to recognize workflow invocations and invoke internal workflow functions. Second, OpenFaas is modified to support function addressing and identification. We maintain a function addressing table in the OpenFaas, which stores the identity, name, namespace, and endpoint of each replica of the function. The difficulty here is that the endpoint of functions can be dynamic, the mapping should also be updated in real-time.

For the workflow development, it contains workflows in control and physical training planes. To build the control plane, scheduler and communicator functions are implemented as stateful functions with a backend in-memory store. To build the physical training plane, we migrate ElasticDL[32] to suit a serverless manner and corresponding processing requirements, like communication. The communicator functions are implemented as gRPC senders and receivers to synchronize the model state. For low-level training, TensorFlow is used.

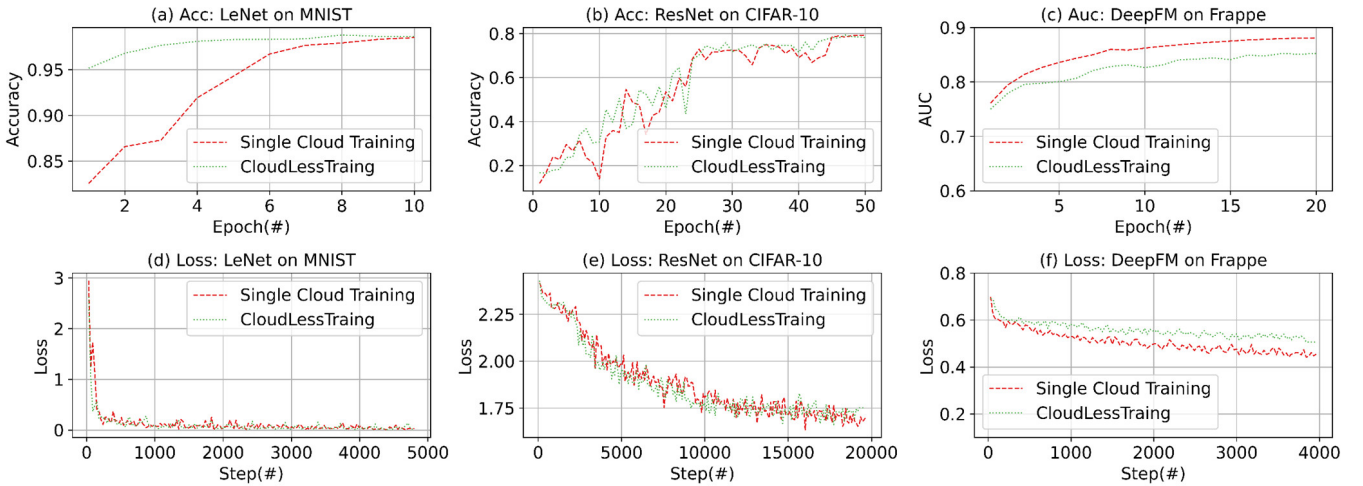

Fig. 7. Accuracy and loss comparison of training LeNet, ResNet, DeepFM with Cloudless-Training in Shanghai and Chongqing regions of Tencent Cloud and trivial PS ML training in Shanghai region of Tencent Cloud

## III. EVALUATION

Cloudless-Training is evaluated for usability, performance of elastic scheduling and communication with 3 AI models in Tencent Cloud, a typical public cloud provider in China, showing that Cloudless-Training can support general ML training in a geo-distributed way, greatly improve resource utilization(e.g., 9.2%-24.0% training cost reduction) and communication performance optimization (e.g., 1.7x training speedup over baseline at most) with model correctness guarantees.

**Environment.** We build 2 serverless clusters in Shanghai and Chongqing regions (locating in the east and west of China respectively) of Tencent Cloud and deploy Cloudless-Training in them. The WAN bandwidth between the 2 clusters is 100Mbps, which is the maximum setting provided by Tencent Cloud. 2 types of CPUs are adopted for training, Intel Xeon Cascade Lake (Cascade) and Intel Xeon Skylake (Sky).

TABLE III. Experimental models and datasets

| Model | Gradient Size | Dataset & Size | Setting |
|---|---|---|---|
| LeNet | 0.4MB | MNIST (60K) | Epoch=10 |
| ResNet | 0.6MB | CIFAR-10 (50k) | Epoch=50 |
| DeepFM | 2.4MB | Frappe (200k) | Epoch=20 |

**ML models.** The experiments are evaluated with 3 AI models, including LeNet[43], ResNet[6], and DeepFM[44], as shown in TABLE III. To reduce the monetary cost of training on Tencent Cloud, the ResNet model used is a variant of ResNet18 that its filters are cut down by a factor of 4. With the models, metrics like training time for specific epochs, accuracy, monetary cost, and communication time on WAN are measured.

## A. Usability

**Specific settings.** To evaluate the usability of Cloudless-Training, we compare its efforts with trivial ML training in a single cloud. For trivial ML training, a PS framework is deployed in Shanghai region with 24 CPU cores (Cascade) and 48GB RAM. For Cloudless-Training, it is equipped with 12 CPU cores (Cascade) and 24GB RAM in each region. In this case, the 2 frameworks have an equal amount of cloud resources, and simple asynchronous SGD is used for model synchronization. Accuracy (or Area Under the Curve - AUC) and loss are measured to evaluate the correctness of model performance.

**Results.** In experiments, all 3 models are successfully trained with Cloudless-Training, as shown in Fig. 7. Cloudless-Training provides a similar trend of accuracy (or AUC) and loss convergence as trivial ML training in all cases. Considering the training accuracy of LeNet, ResNet and DeepFM respectively, Cloudless-Training achieves 0.9864, 0.79, and 0.88, close to 0.9851, 0.78, and 0.84 of trivial ML training. For training loss convergence speed, Cloudless-Training performs competitive results with trivial ML training in both LeNet and ResNet, while slightly slower than trivial ML training in DeepFM.

## B. Performance of Elastic Scheduling

TABLE IV. Resourcing plan settings of elastic scheduling

| ID | Data Distribution Ratio | Baseline Plan (SH/CQ) | Resourcing Plan by Algorithm (SH:CQ) |
|---|---|---|---|
| 1 | 1:1 | Cascade/Sky 12:12 | Cascade/Sky 12:8 |
| 2 | 2:1 | Cascade/Cascade 12:12 | Cascade/Cascade 12:6 |
| 3 | 2:1 | Cascade/Sky 12:12 | Cascade/Sky 12:4 |

**Specific settings.** To evaluate the performance of elastic scheduling strategy, in the 2 regions, Shanghai(SH) and Chongqing(CQ), we adopt different types of CPUs (Cascade, and Sky respectively), both of which can provide 12 cores at most. According to formula (1), the ratio load power of the 2 kinds of resources is about 2:3. The experiments are divided into 3 groups by regulating data distribution ratio, resourcing plan with specific types and core numbers, as shown in TABLE I. All baseline experiments use a greedy strategy to consume all available 24 CPU cores, 12 from each region. The resourcing plan generated by elastic scheduling strategy of Cloudless-Training fully uses CPU cores of the straggler region and cut down the practical allocation of the other region.

**Results.** The resource utilization of geo-distributed training can be significantly improved via elastic scheduling with a comparative or even better accuracy trending, as shown in Fig. 8 and Fig. 9.

First, the resource utilization is greatly improved, and the momentary cost is largely reduced, as shown in Fig. 8. Each pair in Fig. 8(a-c) presents the proportion of training time (including effective execution time and waiting time for stragglers) before and after using the elastic scheduling strategy. In each case, the total training time is basically equal to the baseline, while the waiting time gets reduced. Due to the fluctuation of network, the total time may be slightly higher than baseline sometimes. The waiting time decreases by 46.0%-82.6% for LeNet, 82.3%-94.6% for ResNet, and 6.8%-26.0% for DeepFM. Since DeepFM is more communication-intensive than the other 2 models, its improvement is not as obvious as others. Based on the improvement of resource utilization, the training cost caused by resource consuming while waiting for stragglers is cut down greatly. The training cost is decreased by 13.8%-16.0% for LeNet, 9.2%-15.7% for ResNet, and 13.4%-24.0% for DeepFM. as shown in Fig. 8(d-f).

Second, the accuracy and convergence speed are improved, as shown in Fig. 9. Most of the accuracy curves are slightly higher than the baseline. And the convergence is mostly faster than the baseline and shows fewer vibrations. It is because our scheduling helps balance the training paces among clouds, which decreases the effect of stale gradients. At the same time, it implies that the model trained by our strategy can be more stable.

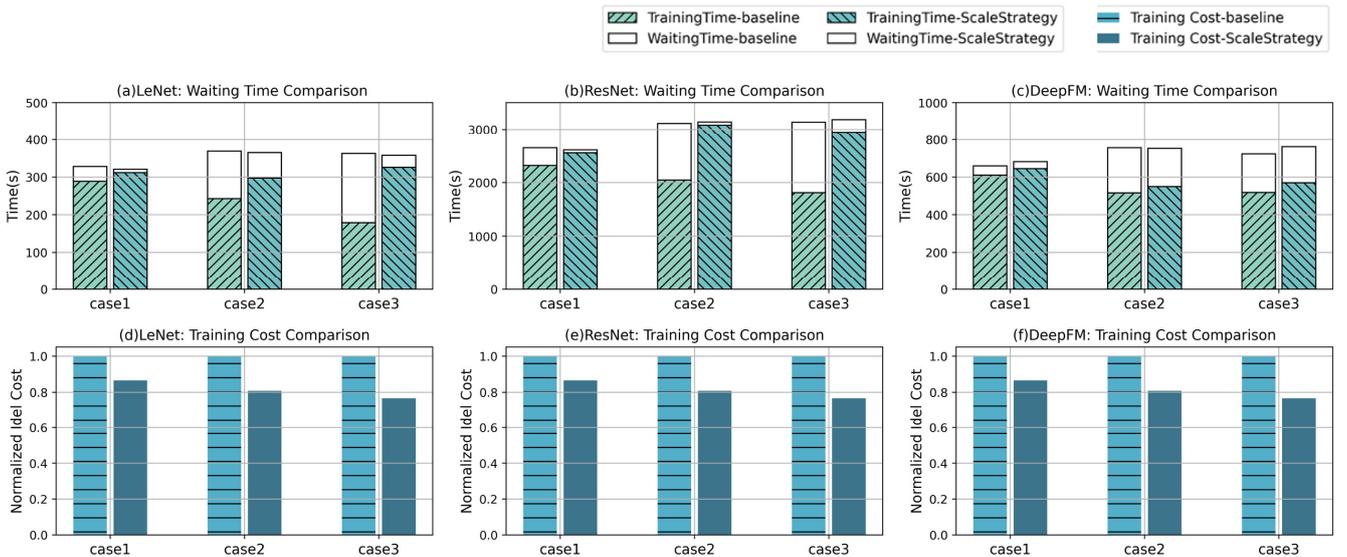

Fig. 8. Training time and cost comparison with and without elastic scheduling in 3 cases with various data distribution (1:1, 1:2, 2:1) and CPU cores allocations in Shanghai and Chongqing regions of Tencent Cloud

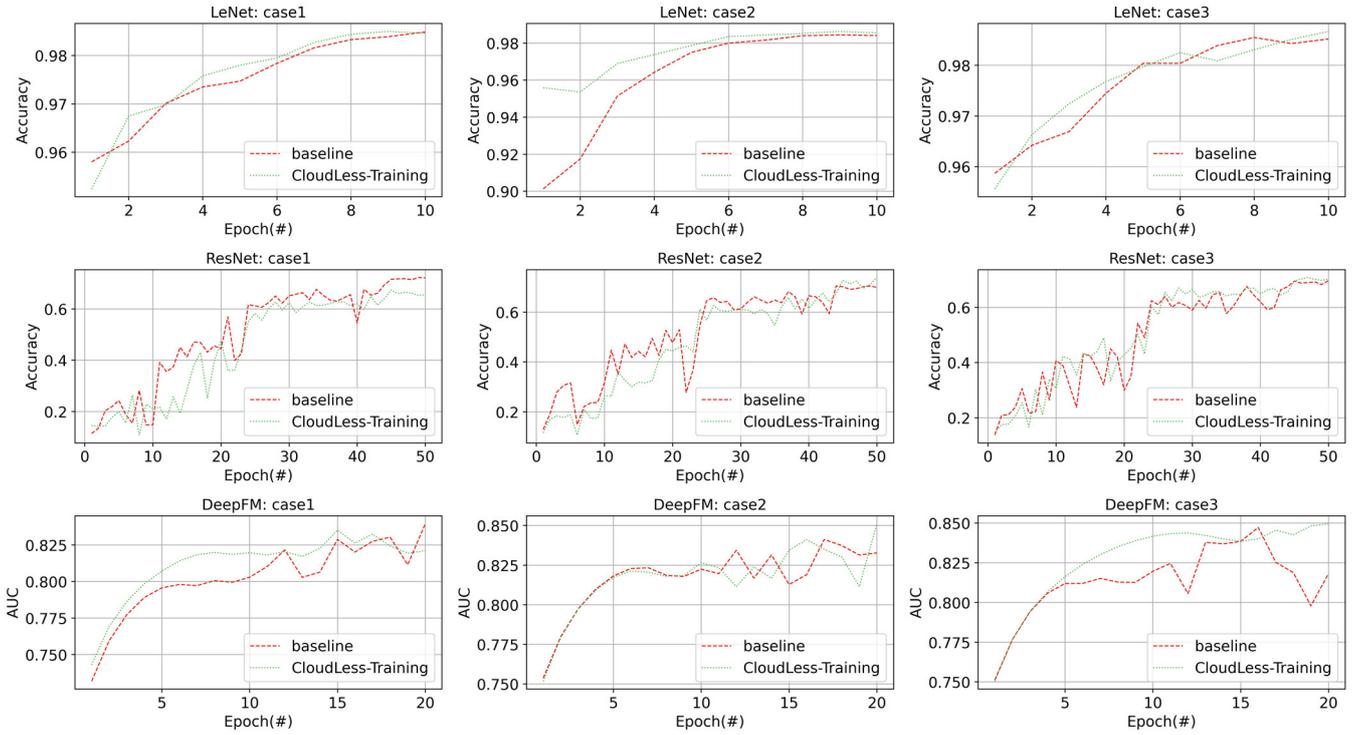

Fig. 9. Accuracy convergence comparison with and without elastic scheduling in 3 cases with various data distribution (1:1, 1:2, 2:1) and CPU cores allocations in Shanghai and Chongqing regions of Tencent Cloud

## C. Performance of Synchronization

**Specific settings.** To evaluate the effect of synchronization optimization, including asynchronous SGD with gradient accumulation (ASGD-GA) and inter-PS model averaging (MA), 4 patterns are observed, including baseline, ASGD-GA, inter-PS model averaging with asynchronous pattern (AMA), and inter-PS model averaging with synchronous pattern (SMA). In baseline, the training in each cloud synchronizes with others via simple asynchronous SGD. The synchronization frequency of baseline is 1 iteration step. It can be seen as a simple multi-regional cloud variant of trivial ML training. For ASGD-GA and AMA, the synchronization frequency is tested with 4 and 8 iteration steps. They are all deployed on Tencent Cloud mentioned above except SMA. For SMA, it is tested in self-hosted cloud clusters in Beijing and Shanghai due to its high monetary cost feature.

**Results.** The performance of geo-distributed training can be significantly improved via the synchronization optimization strategies in Cloudless-Training, as shown in Fig. 10 and Fig. 11.

First, both ASGD-GA and AMA can speed up geo-distributed training effectively, as shown in Fig. 10(a-c). The training speed values of LeNet, ResNet, and DeepFM are increased by 1.2×, 1.2× and 1.7× respectively. The improvement of communication efficiency is mainly from the increasement of synchronization frequency among clouds, from 1 to 4 to 8. It reduces the total communication traffic by several times, and the speedup performance of ASGD-GA and AMA is quite similar. When the synchronization frequency is 4, the communication time of training LeNet, ResNet, and DeepFM with ASGD-GA decreases by 48.3%, 52.7%, and 58.4%, and the time with AMA decreases by 46.0%, 52.7%, and 58.5% than baseline. When it increases to 8, the time decreases by 64.3%, 70.1% and 72.9% with ASGD-GA, and 60.8%, 56.7% and 71.0% with AMA than baseline. Since the fluctuations in WAN, the decline is not as twice as expected in theory.

Second, both ASGD-GA and AMA in Cloudless-Training can guarantee the accuracy convergence of geo-distributed training, as shown in Fig. 10(d-f). In most cases, the trend of the accuracy of ASGD-GA and AMA is similar with the baseline. The delay of gradients or model synchronization may slow down the global synchronization sometimes, but the result in the end can still reach or even exceed the baseline.

Third, SMA performs the best accuracy, as shown in Fig. 11. Comparing baseline, ASGD-GA, AMA, and SMA in a self-hosted environment, the training time of using SMA is much slower than ASGD-GA and AMA, similar to the baseline. However, its accuracy is obviously better than all others. We argue that this can be used to improve training accuracy and is worth further study.

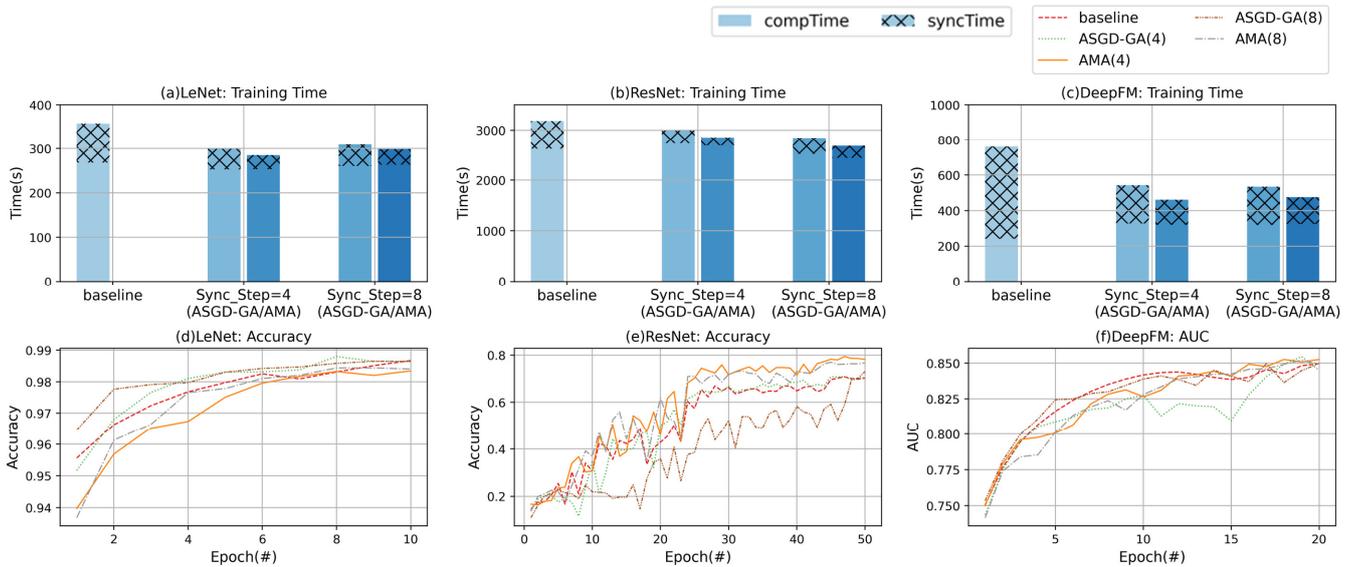

Fig. 10. Training time and accuracy convergence comparison of ResNet with 3 kinds of model synchronization strategies (ASGD, ASGD-GA, AMA) in Shanghai and Chongqing regions of Tencent Cloud

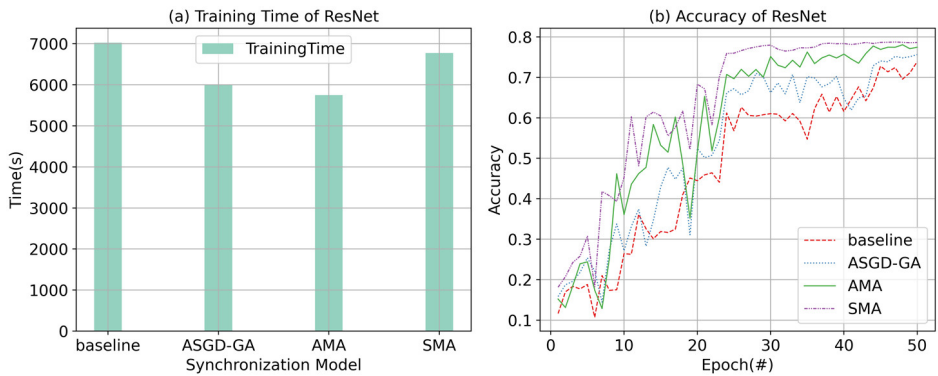

Fig. 11. The training time and accuracy convergence comparison of ResNet with 4 kinds of model synchronization strategies (ASGD, ASGD-GA, AMA, SMA) in a self-hosted environment

## IV. RELATED WORK

**Geo-distributed ML training.** Geo-distributed ML training has attracted wide research interest recently [8][9][10]. It can be seen as a high-level variant of distributed ML training, like Horovod[30] and Kungfu[31]. Some challenges are shared and expanded from distributed to geo-distributed ML training. For example, communication is widely discussed in both distributed and geo-distributed ML training. Studies of geo-distributed ML training mainly extend the mechanisms (e.g., communication) of essential distributed ML training framework and guarantee basic usability. Gaia[8] enables different data centers to communicate at an approaching LAN speed. Nebula-I[9] framework allows heterogeneous cloud clusters to collaboratively train deep learning models. ElasticDL[32] is innate for cross-cloud training as a Kubernetes-native deep learning framework that supports fault-tolerance and elastic scheduling. None of them focus on solving the load balancing problem and exploring various communication optimization strategies like Cloudless-Training. Besides, Cloudless-Training is also designed with serverless architecture.

**Synchronization optimization.** It is widely studied for distributed ML training, and some of them can be migrated into geo-distributed ML training. There are 3 commonly used synchronization strategies in industrial circles: (1) Bulk Synchronous Parallel (BSP), which requires workers to synchronize after each computation[47]. (2) Stale Synchronous Parallel (SSP), which relaxes the limit on synchronization frequency and allows the faster workers to move ahead for a certain number of iterations than the slower workers[48]. (3) Barrierless Asynchronous Parallel(BAP), which allows workers to communicate without any waiting time[49]. Meanwhile, in academic circles, many variants have been proposed. Approximate Synchronous Parallel (ASP), which sends gradients until they reach the significance threshold[8]. Decentralized Parallel Stochastic Gradient Descent (D-PSGD)[46] all workers compute stochastic gradients locally and at the same time average its local model with its neighbors. Asynchronous Decentralized Parallel Stochastic Gradient Descent (AD-PSGD)[45] average its local model with its neighbors with asynchronous mode.

**Serverless computing for ML training**. Functions as a Service (FaaS) has been used to serve distributed ML training, bringing higher resource efficiency, and scaling performance, while it is not widely adopted in geo-distributed ML training yet. Serverless computing frameworks, like Kubeless[36], OpenWhisk[40], OpenFaas[7] and Lambda[41] are used to realize event-driven, elastic computing, attracting practices from many specific application scenarios. For trivial distributed ML training, studies use serverless to improve the elasticity of parallel training, like Cirrus[38], LambdaML[39]. LambdaML presents a FaaS-based training system to determine the cases where FaaS holds sway over IaaS. For geo-distributed ML training with serverless, like Fedless[37], enable federated learning across a large scale of FaaS providers. In this work, Cloudless-Training customizes serverless runtime for supporting geo-distributed ML training scheduling and communication and build a geo-distributed ML training-oriented workflow.

## V. CONCLUSION

In this paper, we present Cloudless-Training, a geo-distributed ML training framework with serverless architecture. It uses 2-layer architecture to simplify logical view of training management for both scheduling and model synchronization in a serverless manner. In terms of scheduling, it provides an elastic scheduling strategy that guarantees load-balanced resource provisioning by modeling heterogeneity of available cloud resources and distribution of pre-existing training datasets. In terms of model synchronization, it builds a basic WAN communication mechanism among clouds based on the WAN communication addressing support, and provides 2 strategies, asynchronous SGD with gradient accumulation (ASGD-GA) and inter-PS model averaging (MA). Experiments show that Cloudless-Training can support general ML training in a geo-distributed way, greatly improve resource utilization (e.g., 9.2%-24.0% training cost reduction) and communication performance (e.g., 1.7x speedup at most of training over baseline) with model correctness guarantee. Thus, we argue that Cloudless-Training is a practical and effective system to improve the efficiency of geo-distributed ML training.